\documentclass[prl,twocolumn,showpacs,superscriptaddress,floatfix]{revtex4}
\pdfoutput=1
\usepackage{graphicx}
\usepackage{dcolumn}
\usepackage{amsmath,amssymb}

\newcommand{\vsd}{\ensuremath{V_{\text{sd}}}}

\newcommand{\vac}{\ensuremath{v_{\text{ac}}}}

\newcommand{\fo}{\ensuremath{f_{0}}}
\newcommand{\qset}{\ensuremath{Q_{\text{SET}}}}
\newcommand{\qo}{\ensuremath{Q_{0}}}
\newcommand{\qt}{\ensuremath{Q_{T}}}

\newcommand{\gd}{\ensuremath{G_{d}}}

\newcommand{\ngt}{\ensuremath{n_{g}}}
\newcommand{\dq}{\ensuremath{\delta q}}

\newcommand{\rn}{\ensuremath{R_{n}}}
\newcommand{\ehz}{\ensuremath{e/\sqrt{\mathrm{Hz}}}}
\newcommand{\aehz}[1]{\ensuremath{#1\:\ehz}}
\newcommand{\e}[1]{\ensuremath{\times 10^{#1}}}
\newcommand{\units}[1]{\ensuremath{\mathrm{#1}}}
\newcommand{\amount}[2]{\ensuremath{#1\:\units{#2}}}
\newcommand{\cp}{\ensuremath{C_{p}}} 
\newcommand{\cg}{\ensuremath{C_{g}}}
\newcommand{\vg}{\ensuremath{V_{g}}}
\newcommand{\ec}{\ensuremath{E_{c}}}
\newcommand{\csig}{\ensuremath{C_{\Sigma}}}

\newcommand{\iv}{$I$-$V$}
\newcommand{\kb}{\ensuremath{k_{B}}}

\newcommand{\sis}{\ensuremath{S_{I}}}
\newcommand{\sqs}{\ensuremath{S_{Q}}}
\newcommand{\zo}{\ensuremath{Z_{0}}}
\newcommand{\wo}{\ensuremath{\omega_{0}}}
\newcommand{\ta}{\ensuremath{T_{\text{HEMT}}}}
\newcommand{\ts}{\ensuremath{T_{\text{SET}}}}
\newcommand{\gs}{\ensuremath{\gamma_{\text{SET}}}}
\newcommand{\tc}{\ensuremath{T_{\text{circ}}}}
\newcommand{\alx}{\ensuremath{{\rm Al/AlO}_{x}}}

\newcommand{\pn}{\ensuremath{P_{n}}}

\newcommand{\pset}{\ensuremath{P_{\text{SET}}}}
\newcommand{\pseti}{\ensuremath{\mathcal{P}_{\text{SET}}}}

\newcommand{\gin}{\ensuremath{\Gamma_{\text{in}}}}
\newcommand{\gout}{\ensuremath{\Gamma_{\text{out}}}}
\newcommand{\kout}{\ensuremath{K_{\text{out}}}}
\newcommand{\eint}{\ensuremath{E_{\text{int}}}}

\begin{document}

\title{Quantum Noise, Effective Temperature, and Damping in a Superconducting Single-Electron Transistor} 

\author{W. W. Xue}\thanks{These authors made equal contributions to this work.}
\affiliation{6127 Wilder Laboratory, Department of Physics and Astronomy,
Dartmouth College, Hanover, NH, 03755 USA}

\author{Z. Ji}\thanks{These authors made equal contributions to this work.}
\affiliation{Department of Physics and Astronomy, Rice University, Houston, TX 77005 USA}

\author{Feng Pan}
\affiliation{6127 Wilder Laboratory, Department of Physics and Astronomy,
Dartmouth College, Hanover, NH, 03755 USA}

\author{Joel Stettenheim}
\affiliation{6127 Wilder Laboratory, Department of Physics and Astronomy,
Dartmouth College, Hanover, NH, 03755 USA}

\author{A. J. Rimberg}\email{ajrimberg@dartmouth.edu}
\affiliation{6127 Wilder Laboratory, Department of Physics and Astronomy,
Dartmouth College, Hanover, NH, 03755 USA}

\begin{abstract}
We have directly measured the quantum noise of a superconducting single-electron transistor (S-SET) embedded in a microwave resonator consisting of a superconducting  $LC$ tank circuit.  Using an effective bath description, we find that the S-SET provides damping of the resonator modes proportional to its differential conductance and has an effective temperature that depends strongly on the S-SET bias conditions.   In the vicinity of a double Cooper pair resonance, when both resonances are red detuned the S-SET effective temperature can be well below both the ambient temperature and the energy scale of the bias voltage.  When blue detuned, the S-SET shows negative differential conductivity, negative damping, and a negative effective temperature.

\end{abstract}
\maketitle

Fluctuations in the current through a nanoscale device are a fertile source of information regarding correlations between charge carriers \cite{Blanter:2000}. In the case of charge and displacement sensors such as single-electron transistors (SETs) these fluctuations are related both to the sensors' ultimate sensitivity and to their backaction on an object under measurement \cite{Mozyrsky:2002,EffTTheory}.  In such cases, one is primarily interested in the noise not near dc, but at frequencies on the order of 1--\amount{100}{GHz}.   In this regime, when the photon energy  becomes comparable to the energy scale of the temperature $\kb T$ or bias voltage $e\vsd$, it becomes important to consider the noise from a quantum perspective \cite{Gardiner:2000}; specifically, one must account for the fact that quantum systems can either emit or absorb energy by considering the unsymmetrized quantum noise spectrum $\sis(\omega)=\int_{-\infty}^{+\infty} dt e^{i\omega t}\langle I(t)I(0)\rangle$ of the current.  In such a treatment, negative frequencies correspond to emission, and positive frequencies to absorption.  

Recently there has been increased theoretical recognition of the importance of such an approach for nanoscale systems \cite{LesovikTheory,Aguado:2000,Gavish:2000} and measurements of high frequency noise in  Cooper pair boxes \cite{Deblock:2003}, Josephson junctions \cite{Billangeon:2006}, superconducting SETs in the Cooper pair tunneling regime \cite{Naaman:2007} and quantum point contacts \cite{Gustavsson:2007} have been performed.  Other measurements have focused on the backaction of a detector such as an S-SET \cite{Naik:2006} or atomic point contact \cite{FlowersJacobs:2007} on a nanomechanical resonator.  Here, we report {\itshape complete and quantitative} measurements of the quantum noise of an S-SET in the vicinity of the double Josephson-quasiparticle (DJQP) transport cycle, which involves tunneling of both Cooper pairs and quasiparticles \cite{JQPExpt}.  The DJQP cycle has drawn recent interest due to the interesting noise properties associated with such resonances \cite{ChoiTheory}, and because an S-SET biased near it is expected to approach the quantum limit for charge detection \cite{Clerk:2002}.  The DJQP cycle has also been the focus of an effective bath description of the quantum noise and backaction of mesoscopic electrical detectors \cite{EffTTheory}.

\begin{figure}[h]
\begin{center}
\includegraphics[width=7.5cm]{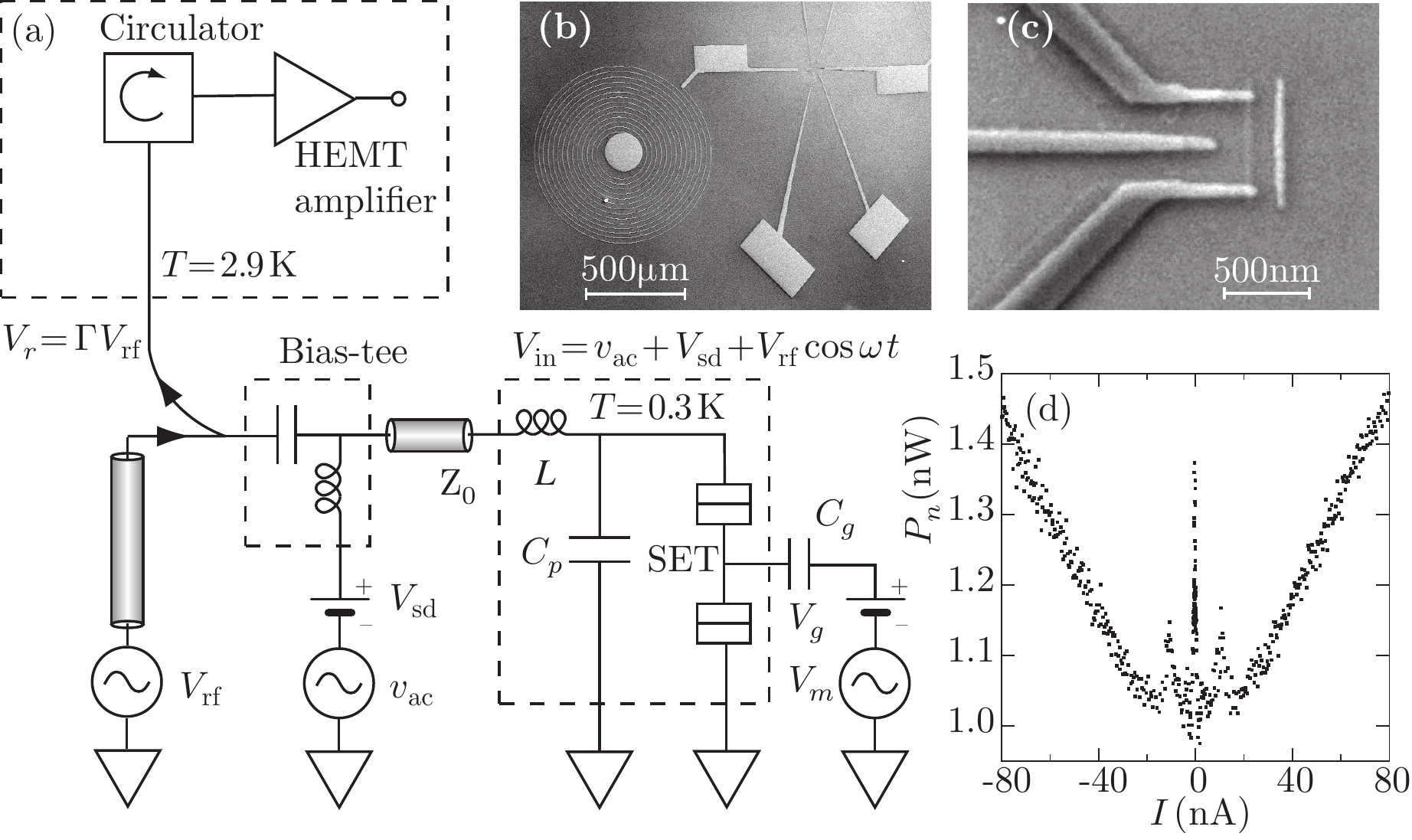} 
\caption{\label{fig1} (a) Measurement circuit, showing the sample, bias-tee, circulator and HEMT amplifier.  Low frequency wiring was filtered, and the input coaxial line included attenuation of \amount{34}{dB}. (b), (c) Electron micrographs of a typical sample, showing the on-chip inductor and an SET\@.  (d) Noise power \pn\ at the output of the amplifier chain versus SET current $I$.   }
\end{center}
\end{figure}

Measurements were performed on \alx\ S-SETs fabricated on a GaAs substrate.  Each S-SET was embedded in an on-chip superconducting microwave resonator consisting of a spiral chip inductor $L$ and parasitic capacitance \cp\ as shown in Fig.~\ref{fig1}(a); such nearly dissipationless resonators \cite{Xue:2007} enable improved transfer of quantum noise power out of the S-SET, making our measurements possible.  The samples were placed in a $^{3}$He refrigerator with a base temperature of \amount{290}{mK}.  A bias-tee attached the samples to both low frequency wiring and, via coaxial lines and a directional coupler, to a circulator and cryogenic HEMT amplifier, both at a temperature of \amount{2.9}{K}.   At room temperature, the HEMT output was  further amplified, allowing either noise measurements or  radio-frequency operation (RF-SET) \cite{Schoelkopf:1998}.   The circulator isolated the sample from the HEMT amplifier, simplifying later analysis \cite{Roschier:2004}.  The resonator is characterized by its loaded $Q$-factor \qt\ given by $\qt^{-1} = \qo^{-1}+\qset^{-1}$ where $\qo=\sqrt{L/\cp}/\zo$ is the external $Q$ of the resonator, $\zo=\amount{50}{\Omega}$ is the feedline impedance, $\qset=\sqrt{\cp/L}/\gd$ is the $Q$ of the S-SET, and \gd\ is its differential conductance.  At the resonant frequency $\wo=1/\sqrt{L\cp}$ the reflection coefficient \gin\ for rf signals incident on the resonator is given by $\gin = \frac{\gd L/\cp - \zo}{\gd L/\cp + \zo}$.  Because the $LC$ circuit is essentially dissipationless, its damping is determined by coupling to the feedline and the S-SET differential conductance \gd.  The simple expressions for \qt, \qset, and \gin\ above therefore allow us to determine $L\approx\amount{167}{nH}$ and $\cp\approx\amount{0.14}{pF}$ following the procedures in Ref.~\onlinecite{Xue:2007}, with $\qo=22$ and $\fo=\wo/2\pi=\amount{1.04}{GHz}$.

To characterize the HEMT amplifier and circulator, we applied a dc current $I$ to the SET and measured the noise power \pn\ at  \wo\ in a bandwidth $\Delta f = \amount{5}{MHz}$ at the output of the  amplifier chain, as shown in Fig.~\ref{fig1}(d).   Here $\pn=A(\kb\ta+|\gin|^{2}\kb\tc+\pset(I))\Delta f$ where \ta\ and \tc\ are the HEMT and circulator noise temperatures, respectively, and $\pset(I)$ is the spectral noise density of the SET \cite{Aassime:2001a,Roschier:2004} referred to the HEMT input.  The total gain $A=\amount{61}{dB}$ of the amplifier chain is determined from the slope of the linear part of \pn\ versus $I$, while $\ta=\amount{9.5}{K}$ (which dominates the amplifier noise) was determined from the intersection of the linear asymptotes. (The circulator's contribution to this measurement was negligible, since $|\gin|^{2}\ll 1$ for large $I$.)  $\tc\approx\amount{2.9}{K}$ was found by measuring \pn\ at $I=0$ and subtracting the HEMT contribution, in excellent agreement with the circulator's physical temperature.  

\begin{figure}[h]
\begin{center}
\includegraphics[width=7.5cm]{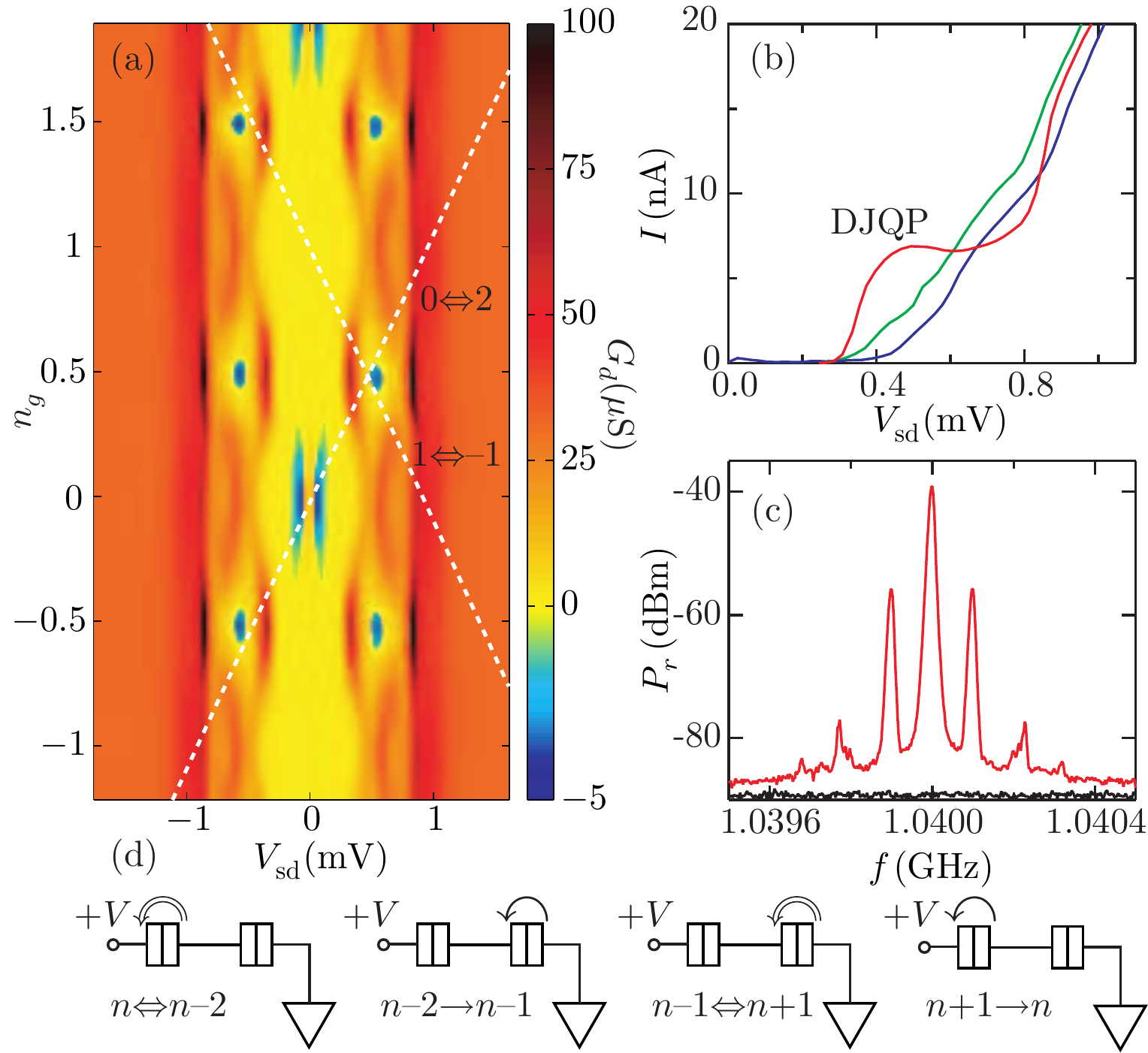} 
\caption{\label{fig2} (a) \gd\ for the S-SET versus \vsd\ and \ngt.  NDC is  visible for \vsd\ and \ngt\ in the vicinity of the supercurrent and the DJQP cycle.   Cooper pair resonances $0\Leftrightarrow 2$ and $1\Leftrightarrow -1$ are shown as the dashed lines.   (b) \iv\ characteristics of the S-SET for various \ngt, emphasizing the presence of NDC on the high-bias side of the DJQP resonance. (c) Amplitude modulated reflected power for a charge modulation of $0.01e$ at \amount{100}{kHz}.  The lower curve is the noise floor of amplifier chain for $I=0$.  (d) DJQP cycle.  The Cooper pair resonance $n\Leftrightarrow n-2$ across one junction is interrupted by quasiparticle tunneling $n-2\rightarrow n-1$ across the other.  The Cooper pair transition $n-1\Leftrightarrow n+1$ is then resonant across the second junction until  quasiparticle tunneling $n+1\rightarrow n$ across the first completes the cycle.}
\end{center}
\end{figure}

We measured five S-SETs, finding similar charge sensitivity \dq\ and $\pset(I)$ for all of them and present results for one, which was fabricated with an ultrathin (\amount{7}{nm}) island (as were two others) \cite{Ferguson:2006}, giving it a larger superconducting gap ($\Delta_{i}=\amount{240}{\mu eV}$) than that of the thicker leads ($\Delta_{l}=\amount{190}{\mu eV}$).  This technique minimizes quasiparticle poisoning \cite{Aumentado:2004} and increases the Josephson coupling $E_{J}=\frac{\Delta_{i}\Delta_{l}}{2(\Delta_{i}+\Delta_{l})} \frac{h/e^{2}}{\rn}\approx\amount{51}{\mu eV}$ where $\rn=\amount{27}{k \Omega}$ is the S-SET's total normal state resistance.  We used standard lock-in techniques to measure the  S-SET differential conductance \gd\ versus dc source-drain bias \vsd\ and island charge number $\ngt=\vg\cg/e$ (\vg\ is the gate voltage) as shown in Fig.~\ref{fig2}(a) by applying a small ac voltage \vac\ (\amount{15}{\mu V} at \amount{11}{Hz}) in addition to \vsd.   \gd\ was  $2e$ periodic in \ngt, showing a supercurrent at $\ngt =0, 2,...$.  Features due to the DJQP cycle \cite{Clerk:2002,Thalakulam:2004} (illustrated schematically in Fig.~\ref{fig2}(d)), occur near the intersection two Cooper pair resonances, one for each junction in the S-SET\@.  In the vicinity of the DJQP cycle the S-SET's quantum noise properties are expected to depend strongly on the SET bias in \vsd\ and \ngt\ with respect to this intersection \cite{Clerk:2002,EffTTheory}, which is manifested by a peak in the S-SET current at $\vsd=2\ec/e$ as in Fig.~\ref{fig2}(b). Here $\ec=e^{2}/2\csig$ is the S-SET charging energy, $\csig=C_{1} + C_{2} +\cg$ is its total capacitance and $C_{1(2)}$ are the junction capacitances. The  DJQP peak location gave $\ec=\amount{237}{\mu eV}$, while fitting Cooper pair resonance lines to the data in Fig.~\ref{fig2}(a) \cite{Thalakulam:2004} allowed us to determine $C_{1(2)} =\amount{174\,(160)}{aF}$ and $\cg=\amount{11}{aF}$ so that the S-SET was very nearly symmetric.  Charge sensitivity measurements as in Fig.~\ref{fig2}(d) gave $\dq\approx\aehz{1.7\e{-6}}$ for operation as an RF-SET\@.

There are several points in the \vsd-\ngt\ plane at which $\gd<0$: here the S-SET exhibits negative differential conductivity (NDC).  These regions are associated with Cooper pair resonances, occurring on the high-bias side of both the supercurrent and DJQP features; NDC near the DJQP is visible in Fig.~\ref{fig2}(b) as decreasing current with increasing bias just past the current maximum.  In terms of the simple picture of resonator damping given above, if $\gd<0$ we expect both $\qset <0$ and $|\gin|>1$.  In physical terms the SET is expected to exhibit {\itshape negative damping}: rather than absorb energy from the resonator, the SET should emit energy into it.  

The regions of NDC have special significance for the S-SET's quantum noise properties.   In the vicinity of the DJQP cycle, if the S-SET is biased below (above) both Cooper pair resonances, the S-SET must absorb (emit) energy for transport to occur; we refer to the S-SET being red (blue) detuned with respect to the DJQP cycle.  Since the S-SETs electromagnetic environment is dominated by the $LC$ resonator, most absorption (emission) will take the form of photon exchange with the tank circuit \cite{Lu:2002}.   Given this asymmetry with respect to emission and absorption, a quantum noise description of the S-SET is appropriate.  Furthermore, since there is a large disparity in time scales between the S-SET tunneling rates (10s of GHz) and the inverse response time of the resonator ($\fo/\qt\approx\amount{50}{MHz}$) we use an effective temperature description \cite{Mozyrsky:2002,EffTTheory} in which the quantum noise of the S-SET at \wo\ is given by
\begin{eqnarray}
\sis(\wo) + \sis(-\wo) & = & 4\kb\ts\cp\gs \label{eq1} \\
\sis(\wo) - \sis(-\wo) & = & 2\hbar\wo\cp\gs. \label{eq2}
\end{eqnarray}
Here, the S-SET is treated as a quantum resistor with conductance \gd\ and effective temperature \ts, a measure of the asymmetry of its quantum noise at \wo, while $\gs=\gd/\cp$ is the rate at which it damps the electromagnetic modes of the resonator \cite{LesovikTheory}.  Note that $\kb\ts$ can be significantly smaller than the energy of either the S-SET's physical temperature $\kb T$ or its bias voltage $e\vsd$.  Furthermore, it can be either positive or negative, as can \gs, depending on whether absorption or emission, respectively, dominates the quantum noise.  

\begin{figure}[h]
\begin{center}
\includegraphics[width=7.5cm]{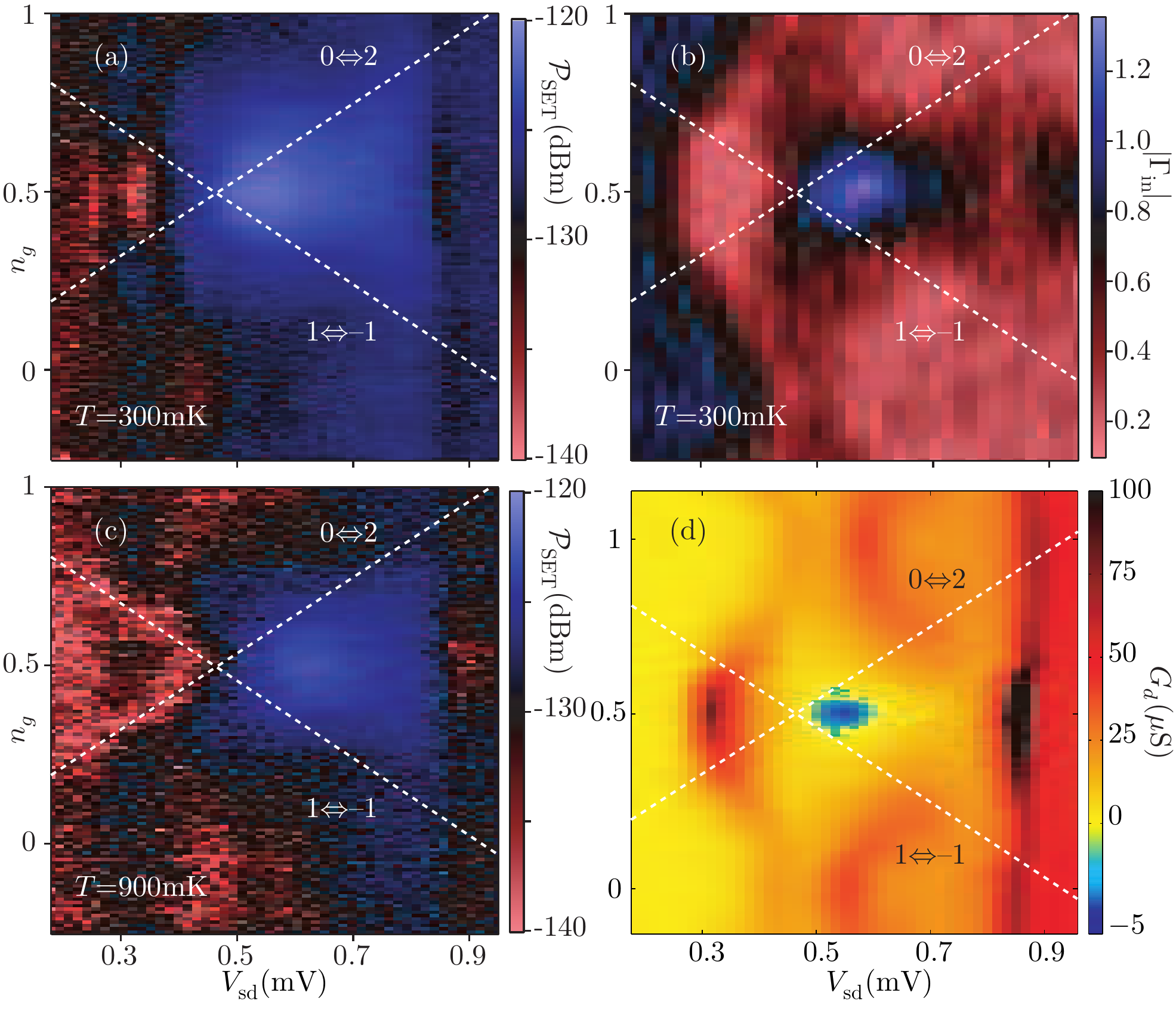} 
\caption{\label{fig3} (a) $\pseti(\vsd,\ngt)$ at \amount{300}{mK}.  Cooper pair resonances are shown by the dashed lines, and the center of the DJQP occurs at their intersection.  Noise is maximal for blue detuning and minimal for red detuning.  (b) $|\gin|(\vsd,\ngt)$ at \amount{300}{mK}.  A small region for which $|\gin|>1$ exists for blue detuning.  (c) At \amount{900}{mK} $\pseti(\vsd,\ngt)$  is smaller in the blue detuned region (in agreement with a lessening of NDC there for higher temperature). The reduction of \pseti\ in the red-detuned region is more pronounced, and tracks nearly exactly the Cooper pair resonance lines.  (d) $\gd (\vsd,\ngt)$ at \amount{300}{mK}.  The region of NDC corresponds nearly exactly to that for which   $|\gin|>1$.  }
\end{center}
\end{figure}

We measured the integrated SET noise $\pseti=\pset\Delta f$, referred to the input of the HEMT at a given \vsd\ and \ngt\ by subtracting the contributions due to the HEMT and circulator. $\pseti(\vsd,\ngt)$ at \amount{300}{mK} in the vicinity of the DJQP resonance is shown in Fig.~\ref{fig3}(a) on a logarithmic scale.  The noise is minimal for red detuning with respect to the DJQP, and maximal for blue detuning.  We can tie these noise characteristics to emission and absorption in the S-SET by measuring the reflection coefficient $|\gin|$ of the tank circuit over the same range of \vsd\ and \ngt, as in Fig.~\ref{fig3}(b).  We applied a very small carrier wave (\amount{-149}{dBm}), measured the reflected power, and after accounting for the HEMT and circulator, computed \gin.  For most values of \vsd\ and \ngt, we found $|\gin|<1$, indicating net absorption by the S-SET\@.  However, when the S-SET is blue detuned, there was a region for which $|\gin|>1$, indicating emission.  Here the S-SET provides {\itshape negative damping}, returning more power to the resonator than is delivered by the rf excitation.  Comparing with the S-SET conductance in the same region as shown in Fig.~\ref{fig3}(d), we see that the region of negative damping corresponds exactly to the region of NDC, in accord with both our expectation based on the forms of \qset\ and \gin, and with the more sophisticated quantum noise viewpoint of (\ref{eq1}) and (\ref{eq2}).  In Fig.~\ref{fig3}(c), we show \pseti\ at \amount{900}{mK}; the general features apparent in Fig.~\ref{fig3}(a) are still visible.

\begin{figure}[h]
\begin{center}
\includegraphics[width=7.5cm]{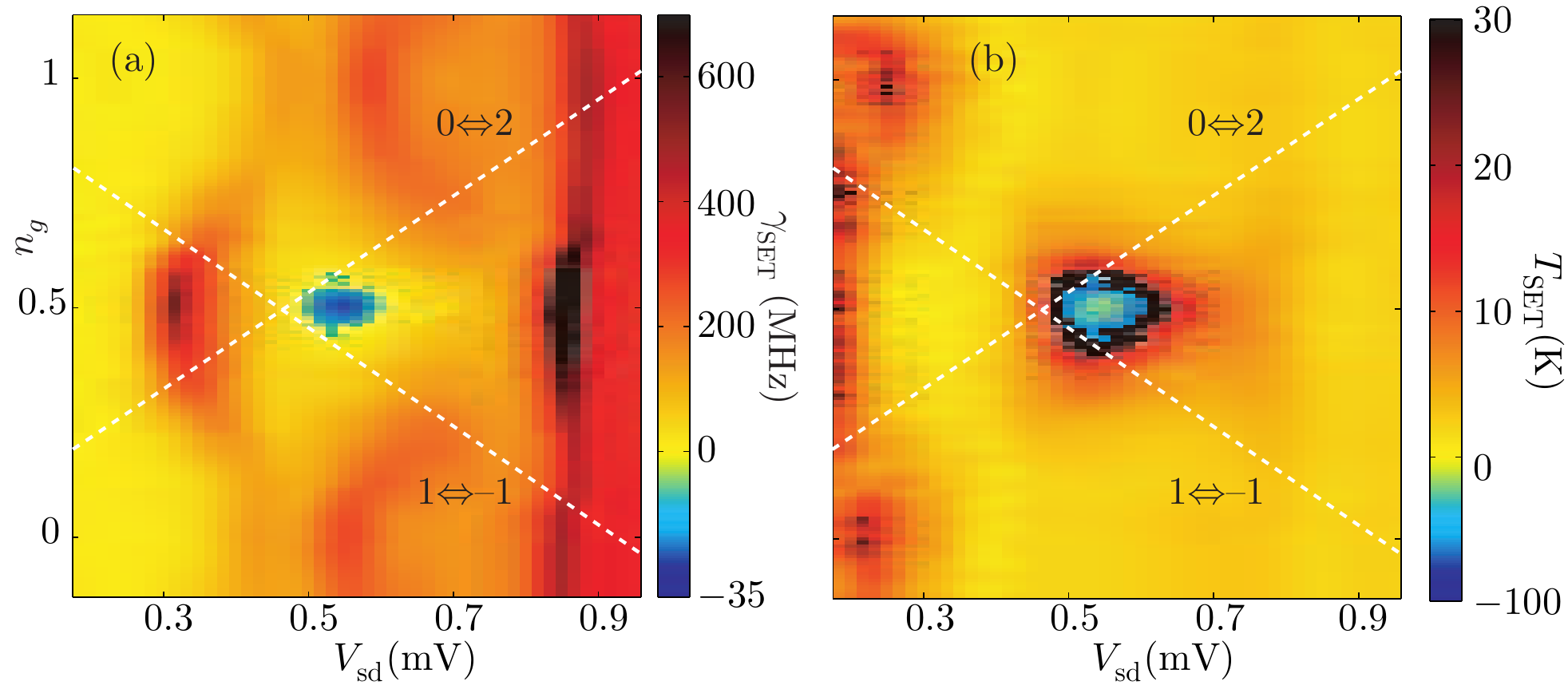} 
\caption{\label{fig4} (a) S-SET damping rate \gs\ and (b) S-SET effective temperature \ts\ at \wo.  Together, these give a complete and quantitative description of the  S-SET quantum noise. }
\end{center}
\end{figure}

Given the excellent correspondence between $|\gin|$ and \gd\ found in our earlier work \cite{Xue:2007} and Fig.~\ref{fig3}, we proceed to use \ts\ and \gs\  to describe the S-SET quantum noise.  We treat the S-SET as a source with available noise power $\kb\ts$ of which some portion $|\gout|^{2}\kb\ts$ is reflected by the resonator while the remainder $\kout\kb\ts$  is delivered to the HEMT\@.  It can be shown that $\pseti=\kout\kb\ts\Delta f=4\qt^{2}\zo\gd\kb\ts\Delta f$,  so that \ts\ can be found from measurements of \pseti, while \gs\ can be determined directly from \gd\ via the relation $\gs=\gd/\cp$.  The resulting values of \gs\ and \ts\ near the DJQP for \amount{300}{mK} are shown in Fig.~\ref{fig4}(a) and (b), respectively, providing a {\itshape complete and quantitative} measurement of the S-SET quantum noise at \wo.  The tendency of the S-SET to either emit or absorb (as measured by \gs) and its degree of asymmetry (as measured by $\ts\propto\frac{\sis(\wo) + \sis(-\wo) }{\sis(\wo) - \sis(-\wo) }$) vary strongly with \vsd\ and \ngt.  For blue detuning (Cooper pairs must give off energy) we observe both {\itshape negative damping} and a {\itshape negative effective temperature}. While \ts\ is large in some areas, for most bias points $\ts\lesssim\amount{1}{K}$ (making it smaller than $e\vsd/\kb$). For red detuning \ts\ can be as low as $\ts\approx\amount{100}{mK}$, less than the ambient temperature.  While theoretical expressions for \gs\ and \ts\ near the DJQP exist \cite{EffTTheory}, they assume capacitive coupling of the S-SET to a resonator rather than our direct electrical connection, and also ignore higher order tunneling processes known \cite{Thalakulam:2004,Naik:2006} to be important for our relatively low-resistance S-SETs.  Additional theoretical work is required for a direct comparison with our experimental results.  Finally, we prefer \ts\ and \gs\ as a description of the S-SET quantum noise over the Fano factor since the latter is due only to fluctuations of the number of tunneling electrons \cite{ChoiTheory}.  In our experiment, variations in \pseti\ arising from electron number fluctuations are indistinguishable from those due to emission/absorption of photons.

Given interest in use of the RF-SET as a potentially quantum limited electrical amplifier \cite{Clerk:2002,Devoret:2000} it is worthwhile to estimate our proximity to this limit.  The uncoupled energy sensitivity $\delta\varepsilon=(\dq)^{2}/2\csig$ ($\delta\varepsilon\approx 1.04\hbar$ for our sample) is frequently used as such an estimate.  However, $\delta\varepsilon$ does not have a lower bound \cite{Devoret:2000}, as a rigorous estimate should.  A better estimate can be obtained when knowledge of both \ts\ and \gs\ is  available.  We imagine coupling the S-SET to some external device such as a quantum dot.  Then the parameter $\chi=\sqrt{4(\eint/e)^{2}\sqs\sis/(\hbar\Delta I)^{2}}$ obeys a strict quantum limit $\chi\geqslant 1$, where \sqs\ is the spectral density of charge fluctuations on the S-SET island,  \eint\ describes its interaction with the measured system, and $\Delta I$ is the change in S-SET current corresponding to a change in the system charge state \cite{Makhlin:2001a}; $\chi^{2}$ is the ratio of the time it takes the S-SET to measure a charge state to the time it takes to dephase it.  Using the voltage fluctuations across the S-SET to estimate $\sqs\approx\frac{1}{4}\kb\ts\csig^{2}/\gd$ we find $\chi\approx\sqrt{2\kb\ts(\dq/e)^2/\gd\hbar^{2}}$ independent of the details of the measured system or its coupling to the S-SET\@.  For typical values $\ts\approx\amount{0.7}{K}$ and $\gd\approx\amount{20}{\mu S}$, and using our measured \dq, we find $\chi\approx2.5$, so that our RF-SET operates in the vicinity of the quantum limit.  

In conclusion, we have directly measured the quantum noise of an S-SET in the vicinity of the DJQP cycle.   We find that the damping of the resonator in which the S-SET is embedded is directly determined by its conductance \gd, allowing us to determine when absorption or emission dominates the quantum noise.  Combining measurements of damping with measurements of total noise \pseti\ from the S-SET allows us to determine its effective temperature \ts, which can be viewed as a measure of the asymmetry of its quantum noise.  \ts\ and the damping \gs\ provide a complete and quantitative description of the quantum noise in the vicinity of the DJQP cycle.  Our measurement technique is very similar to those proposed for measurement of zero-point fluctuations \cite{LesovikTheory,Gavish:2000}. Such measurements, if performed at lower temperatures and possibly higher frequency, could be an interesting area for future investigation, as could be the possibility of producing laser-like instabilities \cite{NegQTheory} if the total quality factor \qt, as opposed to just the S-SET quality factor \qset, can be made negative.  

This work was supported by the ARO under Agreement No.\ W911NF-06-1-0312, by the NSF under Grant No.\ DMR-0454914 and by the NSA, LPS and ARO under Agreement No.\ W911NF-04-1-0389.  We thank M. Blencowe for many helpful conversations and T. J. Gilheart, M. Bal and F. Chen for experimental assistance.


\end{document}